\documentclass[12pt]{article}
\usepackage[dvips]{color}
\usepackage{changebar}
\usepackage{epsfig}
\usepackage{rotating}
\usepackage{amsmath}
\usepackage{hhline}
\usepackage{amssymb}
\usepackage{times}
\usepackage{cite}
\usepackage[left]{lineno}

\newlength{\dinwidth}
\newlength{\dinmargin}
\begin{document}  
\newcommand{\bm}{\boldmath}
\newcommand{\dstar}{$D^\ast$}
\newcommand{\dstarpm}{$D^{\ast\pm}$}
\newcommand{\dstarmu}{$D^\ast\mu$}
\newcommand{\pom}{{I\!\!P}}
\newcommand{\reg}{{I\!\!R}}
\newcommand{\slowpi}{\pi_{\mathit{slow}}}
\newcommand{\fiidiii}{F_2^{D(3)}}
\newcommand{\fiidiiiarg}{\fiidiii\,(\beta,\,Q^2,\,x)}
\newcommand{\n}{1.19\pm 0.06 (stat.) \pm0.07 (syst.)}
\newcommand{\nz}{1.30\pm 0.08 (stat.)^{+0.08}_{-0.14} (syst.)}
\newcommand{\fiidiiiful}{F_2^{D(4)}\,(\beta,\,Q^2,\,x,\,t)}
\newcommand{\fiipom}{\tilde F_2^D}
\newcommand{\ALPHA}{1.10\pm0.03 (stat.) \pm0.04 (syst.)}
\newcommand{\ALPHAZ}{1.15\pm0.04 (stat.)^{+0.04}_{-0.07} (syst.)}
\newcommand{\fiipomarg}{\fiipom\,(\beta,\,Q^2)}
\newcommand{\pomflux}{f_{\pom / p}}
\newcommand{\nxpom}{1.19\pm 0.06 (stat.) \pm0.07 (syst.)}
\newcommand {\gapprox}
   {\raisebox{-0.7ex}{$\stackrel {\textstyle>}{\sim}$}}
\newcommand {\lapprox}
   {\raisebox{-0.7ex}{$\stackrel {\textstyle<}{\sim}$}}
\def\gsim{\,\lower.25ex\hbox{$\scriptstyle\sim$}\kern-1.30ex%
\raise 0.55ex\hbox{$\scriptstyle >$}\,}
\def\lsim{\,\lower.25ex\hbox{$\scriptstyle\sim$}\kern-1.30ex%
\raise 0.55ex\hbox{$\scriptstyle <$}\,}
\newcommand{\pomfluxarg}{f_{\pom / p}\,(x_\pom)}
\newcommand{\dsf}{\mbox{$F_2^{D(3)}$}}
\def\gsim{\,\lower.25ex\hbox{$\scriptstyle\sim$}\kern-1.30ex%
\raise 0.55ex\hbox{$\scriptstyle >$}\,}
\def\lsim{\,\lower.25ex\hbox{$\scriptstyle\sim$}\kern-1.30ex%
\raise 0.55ex\hbox{$\scriptstyle <$}\,}
\newcommand{\dsfva}{\mbox{$F_2^{D(3)}(\beta,Q^2,x_{I\!\!P})$}}
\newcommand{\dsfvb}{\mbox{$F_2^{D(3)}(\beta,Q^2,x)$}}
\newcommand{\dsfpom}{$F_2^{I\!\!P}$}
\newcommand{\gap}{\stackrel{>}{\sim}}
\newcommand{\lap}{\stackrel{<}{\sim}}
\newcommand{\fem}{$F_2^{em}$}
\newcommand{\tsnmp}{$\tilde{\sigma}_{NC}(e^{\mp})$}
\newcommand{\tsnm}{$\tilde{\sigma}_{NC}(e^-)$}
\newcommand{\tsnp}{$\tilde{\sigma}_{NC}(e^+)$}
\newcommand{\st}{$\star$}
\newcommand{\sst}{$\star \star$}
\newcommand{\ssst}{$\star \star \star$}
\newcommand{\sssst}{$\star \star \star \star$}
\newcommand{\tw}{\theta_W}
\newcommand{\sw}{\sin{\theta_W}}
\newcommand{\cw}{\cos{\theta_W}}
\newcommand{\sww}{\sin^2{\theta_W}}
\newcommand{\cww}{\cos^2{\theta_W}}
\newcommand{\trm}{m_{\perp}}
\newcommand{\trp}{p_{\perp}}
\newcommand{\trmm}{m_{\perp}^2}
\newcommand{\trpp}{p_{\perp}^2}
\newcommand{\alp}{\alpha_s}

\newcommand{\alps}{\alpha_s}
\newcommand{\sqrts}{$\sqrt{s}$}
\newcommand{\LO}{$O(\alpha_s^0)$}
\newcommand{\Oa}{$O(\alpha_s)$}
\newcommand{\Oaa}{$O(\alpha_s^2)$}
\newcommand{\PT}{p_{\perp}}
\newcommand{\JPSI}{J/\psi}
\newcommand{\sh}{\hat{s}}
\newcommand{\uh}{\hat{u}}
\newcommand{\MP}{m_{J/\psi}}
\newcommand{\PO}{I\!\!P}
\newcommand{\xbj}{x}
\newcommand{\xpom}{x_{\PO}}
\newcommand{\ttbs}{\char'134}
\newcommand{\xpomlo}{3\times10^{-4}}  
\newcommand{\xpomup}{0.05}  
\newcommand{\dgr}{^\circ}
\newcommand{\pbarnt}{\,\mbox{{\rm pb$^{-1}$}}}
\newcommand{\gev}{\,\mbox{GeV}}
\newcommand{\WBoson}{\mbox{$W$}}
\newcommand{\fbarn}{\,\mbox{{\rm fb}}}
\newcommand{\fbarnt}{\,\mbox{{\rm fb$^{-1}$}}}
%
%
\newcommand{\qsq}{\ensuremath{Q^2} }
\newcommand{\gevsq}{\ensuremath{\mathrm{GeV}^2} }
\newcommand{\et}{\ensuremath{E_t^*} }
\newcommand{\rap}{\ensuremath{\eta^*} }
\newcommand{\gp}{\ensuremath{\gamma^*}p }
\newcommand{\dsiget}{\ensuremath{{\rm d}\sigma_{ep}/{\rm d}E_t^*} }
\newcommand{\dsigrap}{\ensuremath{{\rm d}\sigma_{ep}/{\rm d}\eta^*} }
\def\Journal#1#2#3#4{{#1} {\bf #2} (#3) #4}
\def\NCA{\em Nuovo Cimento}
\def\NIM{\em Nucl. Instrum. Methods}
\def\NIMA{{\em Nucl. Instrum. Methods} {\bf A}}
\def\NPB{{\em Nucl. Phys.}   {\bf B}}
\def\PLB{{\em Phys. Lett.}   {\bf B}}
\def\PRL{\em Phys. Rev. Lett.}
\def\PRD{{\em Phys. Rev.}    {\bf D}}
\def\ZPC{{\em Z. Phys.}      {\bf C}}
\def\EJC{{\em Eur. Phys. J.} {\bf C}}
\def\CPC{\em Comp. Phys. Commun.}

\begin{titlepage}

\noindent

\begin{flushleft}
DESY 05-040 \hfill ISSN 0418-9833 \\
March 2005
\end{flushleft}

\vspace{2cm}

\begin{center}
\begin{Large}

{\boldmath \bf
 Measurement of Charm and Beauty  Photoproduction at HERA
      using \boldmath{\dstarmu}\ Correlations}
\vspace{2cm}

H1 Collaboration

\end{Large}
\end{center}

\vspace{2cm}

\begin{abstract}

\noindent
A measurement of charm and beauty photoproduction at the electron proton collider HERA is presented 
based on the simultaneous detection of a \dstarpm\ meson and a muon.
The correlation between  the \dstar\ meson and the muon serves to separate
the charm and beauty contributions and the analysis provides comparable
sensitivity to both.  The total and differential experimental cross sections
are compared to LO and NLO QCD calculations. The measured charm  cross section
is in good agreement with QCD predictions including higher order effects while
the beauty cross section is higher. 
\end{abstract}

\vspace{1.5cm}

\begin{center}
To be submitted to Phys. Lett. B
\end{center}

\end{titlepage}

%
%

\begin{flushleft}

A.~Aktas$^{10}$,               
V.~Andreev$^{26}$,             
T.~Anthonis$^{4}$,             
S.~Aplin$^{10}$,               
A.~Asmone$^{34}$,              
A.~Astvatsatourov$^{4}$,       
A.~Babaev$^{25}$,              
S.~Backovic$^{31}$,            
J.~B\"ahr$^{39}$,              
A.~Baghdasaryan$^{38}$
% ...

\bigskip{\it
% ...
$ ^{4}$ Inter-University Institute for High Energies ULB-VUB, Brussels;
          Universiteit Antwerpen, Antwerpen; Belgium$^{ c}$ \\
% ...
$ ^{10}$ DESY, Hamburg, Germany \\
% ...
$ ^{25}$ Institute for Theoretical and Experimental Physics,
          Moscow, Russia$^{ l}$ \\
 $ ^{26}$ Lebedev Physical Institute, Moscow, Russia$^{ e}$ \\
% ...
$ ^{31}$ Faculty of Science, University of Montenegro,
          Podgorica, Serbia and Montenegro \\
% ...
$ ^{34}$ Dipartimento di Fisica Universit\`a di Roma Tre
          and INFN Roma~3, Roma, Italy \\
% ...
$ ^{38}$ Yerevan Physics Institute, Yerevan, Armenia \\
 $ ^{39}$ DESY, Zeuthen, Germany \\
% ...
$ ^c$ Supported by FNRS-FWO-Vlaanderen, IISN-IIKW and IWT
      and  by Interuniversity
Attraction Poles Programme,
      Belgian Science Policy \\
% ...
$ ^e$ Supported by the Deutsche Forschungsgemeinschaft \\
% ...
$ ^l$ Partially Supported by Russian Foundation
      for Basic Research, grant    no. 00-15-96584 \\
}
\end{flushleft}

\newpage
\section{Introduction}
\noindent
At the electron proton collider HERA, heavy quarks are predominantly produced via
photon-gluon fusion, $\gamma g\rightarrow c\bar{c}\ \mbox{or} 
\ b\bar{b}$, where the 
photon is emitted from the incoming lepton and the gluon from the proton.
 The production cross sections 
are largest for photoproduction, i.e. for photons with virtuality $Q^2\simeq 0$.
The light quarks $u$, $d$ and $s$ are produced 
much more copiously than $c$ and $b$, and beauty production is suppressed 
by a factor of approximately 200 compared to charm. 
Charm and beauty measurements performed at HERA so far
relied on the tagging of only one heavy quark in each event.
While the charm
measurements~\cite{Aktas:2004ka,Adloff:2001zj,Adloff:1998vb,Aid:1996hj,Chekanov:2003rb,Breitweg:1999ad,Breitweg:1997mj,Breitweg:1998yt,Breitweg:1997gc,Chekanov:2003bu,Breitweg:2000qi}
were mostly based on the reconstruction of $D$\ mesons, the beauty
measurements~\cite{Aktas:2004az,ourbtomupaper,Adloff:1999nr,Chekanov:2003,Chekanov:2004,Breitweg:2000nz}
used semi-leptonic decays or lifetime signatures or both. 
Here an analysis is presented, 
where  in a large fraction of events {\em both} heavy quarks are tagged
using a \dstarpm\ meson and a muon as signatures. 
The correlations between the direction of the muon with respect to the \dstarpm\
and their electric charges are used to separate charm and beauty contributions.
Total cross sections are measured separately for the processes
$ep \rightarrow ec\bar{c} X \rightarrow eD^{*} \mu X'$ and 
$ep \rightarrow eb\bar{b} X \rightarrow eD^{*} \mu X'$ in the visible 
kinematic region, while differential 
cross sections are derived for combined samples of  $c\bar{c}$ and $b\bar{b}$ events. 
The measurements, which are based on an integrated 
luminosity of $\mathcal{L}=89\;\mbox{pb}^{-1}$, are 
compared to leading  order (LO) and next-to-leading order (NLO)
perturbative QCD (pQCD) calculations. 

This measurement extends to significantly lower centre-of-mass energies of the $b\bar{b}$ system
than previous measurements of beauty cross sections at HERA.
The simultaneous detection of the \dstar\ meson and the muon 
makes possible new tests of higher order QCD effects.
For instance, in the photon-gluon rest frame the angle between the heavy quarks
is $180^{\circ}$ at leading order, but at next-to-leading order it can differ 
significantly from this value due to hard gluon radiation.
Furthermore, the \dstarmu\ pair is expected to be sensitive to a possible 
transverse momentum $k_t$ of the gluons entering the quark pair production process. 

\section{Separation of Charm and Beauty}\label{corsec} 
The separation of charm and beauty contributions exploits the charge and azimuthal 
angle\footnote{The coordinate system is defined with the $z$-axis
pointing in the proton beam direction and $x$ ($y$) pointing
in the horizontal (vertical) direction. The azimuthal angle $\Phi$ is measured in the $x$-$y$ plane
and the polar angle $\theta$ with respect to the $z$ direction.} correlations
of the $D^{*}$ meson and the muon. The azimuthal 
angle difference $\Delta\Phi$ between the $D^{*}$ and the muon and their respective electric 
charges $Q(D^\ast)$ and $Q(\mu)$  are used to define
four `correlation regions' I--IV. For $Q(D^{*}) = Q(\mu)$ regions~I and II
cover $\Delta \Phi < 90^\circ$ and $\Delta \Phi > 90^\circ $, respectively.
Regions III and IV are defined correspondingly for $Q(D^{*}) \neq Q(\mu)$.

\begin{figure}[b!]\centering
\unitlength1cm
\epsfig{file=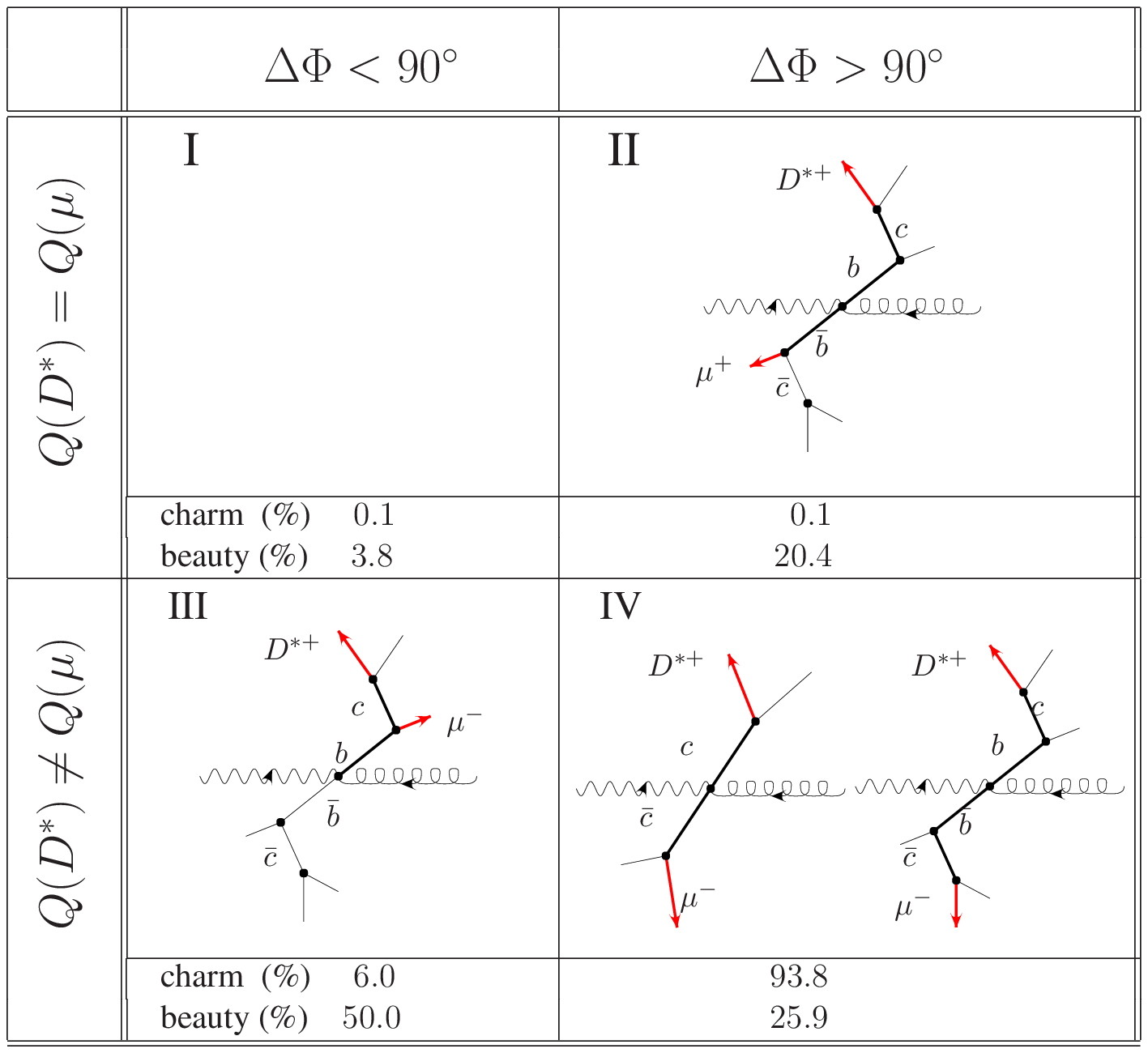,width=14.2cm,clip=}
\caption{\em Definition of the correlation regions in terms of $\Delta \Phi$
             and the relative charges of the \dstar\ meson and the muon. The sketches
             illustrate these correlations in $c\bar{c}$ and $b\bar{b}$ quark
             decays to \dstarmu. 
             The numbers represent the relative distribution over the four correlation 
             regions of charm and beauty events that satisfy the cuts given in 
             section~\ref{dataanev}, as obtained from the
             PYTHIA simulation without detector effects.}
 \label{sketch}
\end{figure}

The four regions are populated differently by charm and beauty events as
is illustrated in figure~\ref{sketch}. Neglecting any transverse momenta of the
photon and the gluon, the fusion process $\gamma g\rightarrow c\bar{c}\ \mbox{or} 
\ b\bar{b}$ leads to a back-to-back configuration of the two heavy quarks.
Approximating the directions of the \dstarpm\ meson and the
muon with those of the quark and antiquark and selecting opposite charges,
$c\bar{c}$ pairs populate correlation region IV. 
In contrast, beauty events populate regions II, III and IV, depending
on whether the muon originates from the same $b$ quark as the \dstar\ or from
the opposite $\bar{b}$. 
If the muon originates from the same $b$ quark as the \dstar\ meson, the events
lie in region~III. For muons coming from the $\bar{b}$ opposite to the \dstar\ meson, the 
direct decay populates region II, while the cascade process
$\bar{b}\to\bar{c}\to\mu$ populates region IV. Region IV hence receives
contributions from both $c\bar{c}$ and $b\bar{b}$ events and region I stays empty.

The azimuthal angle correlations are smeared by fragmentation and  
semileptonic decay processes and by higher order QCD effects such as gluon radiation 
and any initial transverse momentum of the gluon. Processes such as  
heavy quark decays via $\tau$ leptons conserve the charge correlation, which is not the case for 
$B^{0}$-$\bar{B}^{0}$ mixing and e.g. the decays 
$b\rightarrow cW^-; W^-\rightarrow\bar{c}s$.
According to the PYTHIA Monte Carlo simulation~\cite{Sjostrand:1993yb}, 
which takes into account these smearing effects, 
the relative population of the four regions 
is as given by the numbers in figure~\ref{sketch}. 
These numbers apply in the analysed kinematic region 
defined in section~\ref{dataanev} and do not include detector effects. 
Since the population of the four correlation regions is very different for 
$b\bar{b}$ and $c\bar{c}$, it can be used for the separation of these two components.  

\section{QCD Calculations and Monte Carlo Simulations}
The Monte Carlo simulations 
PYTHIA~\cite{Sjostrand:1993yb} and CASCADE~\cite{Jung:2000hk} are used  
for the description of the signal and background distributions in the
separation of charm and beauty, for the determination of efficiencies and acceptances
and for systematic studies. 
Their predictions are also compared with the measured cross sections.
In PYTHIA and CASCADE leading order matrix elements which take into account
the mass of the heavy quark
are implemented and parton showers in the 
initial and final state are included to approximate higher orders (LO-ME+PS).
The parton evolution in PYTHIA uses the DGLAP equations~\cite{Gribov:ri}.
In addition to the direct process, a resolved photon component is generated in PYTHIA
where the photon fluctuates into a hadronic state acting as a source of partons, one of
which participates in the hard interaction. This component is dominated by 
heavy flavour excitation processes\cite{Norrbin:2000zc}, where the
heavy quark is a constituent of the resolved photon. In the PYTHIA calculation of
heavy flavour excitation, in which quark masses are neglected, the 
contribution of excitation to the total charm cross section in the analysed kinematic region
is found to be 41\%, while it is 23\% for beauty. 
In comparison to this component the contribution of the
resolved component due to light quarks or gluons in the photon can be 
neglected in the present analysis, as can the heavy flavour component of the proton.

CASCADE contains an implementation of the CCFM~\cite{ccfm} evolution equation
for the initial state parton shower. The $\gamma g \rightarrow 
c\bar{c}$ or  $b\bar{b}$ is implemented using off-shell matrix elements convoluted 
with $k_t$ unintegrated proton parton distributions.
PYTHIA and CASCADE use the JETSET program as implemented in PYTHIA 
for the hadronisation (via the Peterson fragmentation function~\cite{peterson}) and  
for the decay of beauty and charm quarks.
In order to correct for detector effects, the 
generated events are passed through a detailed
simulation of the detector response based on the
GEANT program~\cite{Brun:1987ma} and the same reconstruction software as used for the data.
\begin{table}[t!]
\begin{center}
\begin{tabular}{|ll|ll|}\hline
\multicolumn{4}{|c|}{Fragmentation Fractions [\%]}\\ \hline
$c\rightarrow D^*$ & $23.5\pm 0.7$ &
$c\rightarrow \mu$ & $9.8\pm 0.5$ \\
$b\rightarrow D^*$ & $17.3\pm 1.6$&
$b\rightarrow \mu$ & $10.95\pm 0.27$\\
$b\rightarrow c\rightarrow\mu$ & $10.03\pm 0.64$ &
$b\rightarrow D^*\mu$ & $2.75\pm 0.19$\\\hline
\end{tabular}
\caption{\em Fragmentation fractions~\cite{opalhf,PDG2002} used in the QCD calculations of the cross
  sections. The $b\rightarrow c\rightarrow\mu$ fraction also contains
  the $b\rightarrow c\bar{c}s$ decay and the $\tau$ contributions.}
\label{tabbr}
\end{center}
\begin{center}
{\footnotesize
\begin{tabular}{|l@{\hspace{.4cm}}|c@{\hspace{.4cm}}c@{\hspace{.4cm}}c@{\hspace{.4cm}}c@{\hspace{.4cm}}|}\hline
  & {\bf PYTHIA}
  & {\bf CASCADE}
  & {\bf FMNR (NLO)}& {\bf FMNR (LO)}
  \rule[-2mm]{0mm}{5mm}
  \\
\hline
Version       &   6.1
              &   1.2007 
              & &
              \\
\hline
Proton PDF
              &  CTEQ5L~\cite{cteq5l}
              &  J2003~\cite{Jung:2000hk}
              &  CTEQ5M~\cite{cteq5l}&CTEQ5L~\cite{cteq5l}
              \\
 
Photon PDF
              &  GRV-G LO~\cite{grvg}
              &  {--}
              &  GRV-G HO~\cite{grvg}& GRV-G LO~\cite{grvg}
              \\
\hline
Renorm. scale    $\mu_r^2$ \rule[-1mm]{0mm}{6mm}
               &   $m_q^2 + p_{Tq}^2$
               &   $4m_q^2 + p_{Tq}^2$
               & \multicolumn{2}{c|}  {$m_q^2 + p_{Tq\bar{q}}^2$} 
               \\
Factor. scale $\mu_f^2$ \rule[-3mm]{0mm}{6mm}
               & $m_q^2 + p_{Tq}^2$
               & $\hat{s} + Q_T^2$
               & \multicolumn{2}{c|}  {$m_b^2 + p_{Tb\bar{b}}^2$,\ 4($m_c^2 + p_{Tc\bar{c}}^2$)} 
               \\
\hline
$m_b \ [ {\rm GeV} ]$           \rule[-1mm]{0mm}{6mm}
               & 4.8
               & 4.8
               & \multicolumn{2}{c|}  {4.75}
               \\
$m_c \ [ {\rm GeV} ]$
               & 1.5
               & 1.5
               & \multicolumn{2}{c|}  {1.5}

               \\
\hline
Peterson $\,\epsilon_b$  \rule[-1mm]{0mm}{6mm}
               & 0.008
               & 0.008
               & 0.0033 (0.42 for $b\rightarrow D^\ast$ )&0.0069
               \\
Peterson $\,\epsilon_c$
               & 0.078
               & 0.078
               & 0.035 & 0.058              \\
\hline
\end{tabular}
\caption{\em 
Parameters used in the Monte Carlo and NLO programs. 
The FMNR calculations are performed in the $\overline{MS}$ scheme using the
default values of  $\Lambda_{QCD}$ for the parton density functions.
$\mu_r$ and $\mu_f$ denote the renormalisation and factorisation
scales, $p^2_{Tq\bar{q}}$  the average of the squares of the 
transverse momenta
of the two heavy quarks, $m_q$ the heavy quark masses,
$\hat{s}$ the centre-of-mass energy squared, $Q_T^2$ the
transverse momentum squared of the heavy quark system, 
$p_{Tq}$ the transverse momentum of a heavy quark and
$\epsilon_q$ the Peterson fragmentation parameters~\cite{Nason:1999zj}.
For the $B^0$-$\bar{B}^0$ mixing values of $x_d\equiv\Delta m_{B^0_d}/\Gamma_{B^0_d}=0.73$ and  
$x_s=18$ \cite{PDG2002} are used.}
\label{tab=MC}}
\end{center}
\end{table}

The measured cross sections  are also compared with 
NLO pQCD calculations in the massive scheme~\cite{massive} using the
program FMNR~\cite{Frixione:1994dv}. 
These calculations are expected to give reliable results in the kinematic region considered
here, where the transverse momentum of the heavy quark is of the same order 
of magnitude as its mass. 
The calculations are available for both the 
direct and resolved photon processes.
However, in contrast to the PYTHIA program, heavy flavour excitation is not explicitly  
included in the resolved part of the FMNR program. The contributions of the resolved
light quark and 
gluon components are found to be small in FMNR 
($<3$\% for charm and $<6$\% for beauty in the analysed kinematic region) 
and are neglected. 

The original FMNR program is extended to include the effects of the
hadronisation of $c$ and $b$ quarks and their semileptonic decays in order to make 
comparisons with the measured cross sections in the experimentally
accessible kinematical region.
The heavy quark is `hadronised' by rescaling the three momentum of the quark
using the Peterson fragmentation function.
For the decay into muons the momentum spectrum is implemented as obtained 
from JETSET. In the case of beauty quarks, the direct decays of
$b$-flavoured hadrons into muons are taken into account as are the decays
via a charm quark, $b\rightarrow c\rightarrow \mu$.
When the \dstar\ meson and the muon  originate from the same quark,
the angular and momentum correlations are implemented as in JETSET. 
The measured fragmentation fractions~\cite{opalhf,PDG2002} for $c$ and $b$ quarks 
given in table~\ref{tabbr} are used for the calculation of the cross sections.
The important parameters of the  pQCD programs used in this analysis are
summarised in table~\ref{tab=MC}. 

\section{Data Analysis}\label{dataan}
The data were collected with the H1 detector \cite{Abt:1997xv,Appuhn:1996na} at 
HERA during the years 1997 to 2000 
and correspond to an integrated luminosity of $\mathcal{L}=89\;\mbox{pb}^{-1}$. 
The largest part of the luminosity (80\%) was collected 
at a centre-of-mass energy of $\sqrt s\approx 320\,$GeV, the beam energies being
27.6 GeV and 920 GeV for electrons\footnote{HERA has been operated with electron and
 positron beams. These periods will not be distinguished in this analysis.} and protons, 
respectively. The remaining
20\% of the luminosity was taken at $\sqrt s\approx 300\,$GeV (proton energy 820 GeV).

\subsection{Event Selection}\label{dataanev}
A detailed account of this analysis can be found in \cite{Wagner:2004ed}.
Events with at least one reconstructed $D^{*}$ and at least one muon are selected;
multiple $D^{*}$ or muon combinations are treated as separate events.
The $D^{*}$ is reconstructed via the decay channel\footnote{Charge
conjugate states are always implicitly included.} $D^{*+}\rightarrow
D^{0}\pi_{s}^{+}\rightarrow K^{-}\pi^{+}\pi_{s}^{+}$ (branching ratio 
$(2.59\pm0.06)\%$~\cite{PDG2002}), where $\pi_{s}$ 
refers to the low momentum $\pi$ in the decay. 
The decay particles of the \dstar\ meson are reconstructed in the central
tracking detector ($20^\circ\leq\theta\le160^\circ$)
without particle identification.
Muons are identified by reconstructing track segments in 
the instrumented iron return yoke of the solenoidal
magnet. These are linked to tracks in the central tracking detector. 
In order to ensure good detector acceptance, cuts on the transverse momentum 
$p_{T}$ with respect to the proton direction  and the pseudorapidity $\eta =
-\ln \; \tan (\theta/2)$ are applied for the $D^{*}$ meson and the muon 
in the laboratory frame (see table~\ref{DsMuCuts}). 

Photoproduction events are selected by demanding the absence of any signals 
for the scattered electron, restricting the accepted range of negative
four-momentum transfer squared $Q^2$ to be below 1 GeV$^2$.
A cut on the inelasticity \mbox{0.05 $<y<$ 0.75} is applied, where 
$y={P\cdot  q}/{P\cdot k}$ ($q$, $k$ and $P$ are the four vectors of the exchanged
photon, incoming electron and proton, respectively). 
The variable $y$ is reconstructed from the measured hadronic final state using
the Jacquet-Blondel method~\cite{jb}.
The events are triggered by fast signals from the central tracking and muon detectors.
The analysed `visible' kinematic region of the measurement is defined 
in table~\ref{DsMuCuts}. 
\begin{table}[h!] 
\begin{center}
\begin{tabular}{|p{6.2cm}|c|}\hline
Selection of $D^*\rightarrow D^0\pi_s\rightarrow K\pi\pi_s$ &                  
                             \multicolumn{1}{|c|}{$p_T(K),p_T(\pi)>0.4\;\mbox{GeV}$}\\
& \multicolumn{1}{|c|}{$p_T(\pi_s)>0.12\;\mbox{GeV}$} \\

& \multicolumn{1}{|c|}{$\mid m_{K\pi}-m_{D^0}\mid <0.080\;\mbox{GeV}$} \\
 & \multicolumn{1}{|c|}{$\Delta M = m_{K\pi\pi_s}-m_{K\pi}< 0.1685
 \;\mbox{GeV}$} \\
\hline
 Visible kinematic region & \multicolumn{1}{|c|}{$p_T(D^*)>1.5\;\mbox{GeV}$} \\
 & \multicolumn{1}{|c|}{$\mid \eta(D^*) \mid <1.5$} \\
 & \multicolumn{1}{|c|}{$p(\mu)>2\;\mbox{GeV}$} \\
 & \multicolumn{1}{|c|}{$\mid \eta(\mu)\mid <1.735$}\\[0.1cm]
 & $0.05<y<0.75$ \\
 & $Q^2<1\;\mbox{GeV}^2$ \\[0.1cm] \hline
\end{tabular}
\caption{\em The \dstar selection cuts and definition of the visible kinematic region.}
\label{DsMuCuts}
\end{center}
\end{table}
\subsection{Fit Procedure}
\label{secfit}
\begin{figure}[h]
  \begin{center}
    \unitlength1cm
 \begin{tabular}{c}
{\epsfig{file=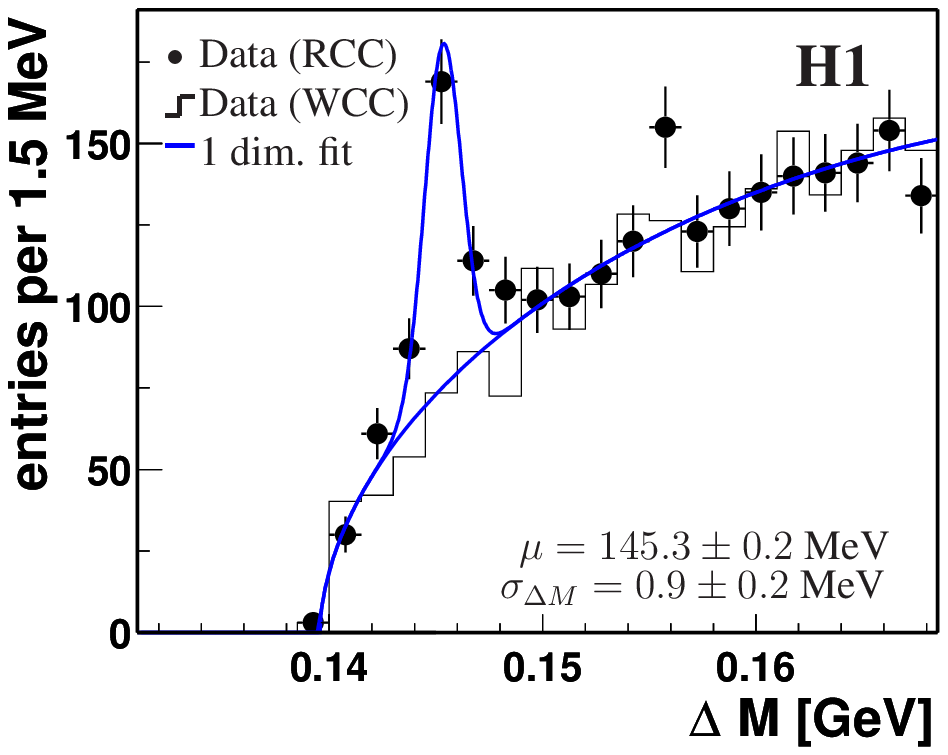,width=9.2cm,angle=0}}  \\
\hspace*{-1cm}{\epsfig{file=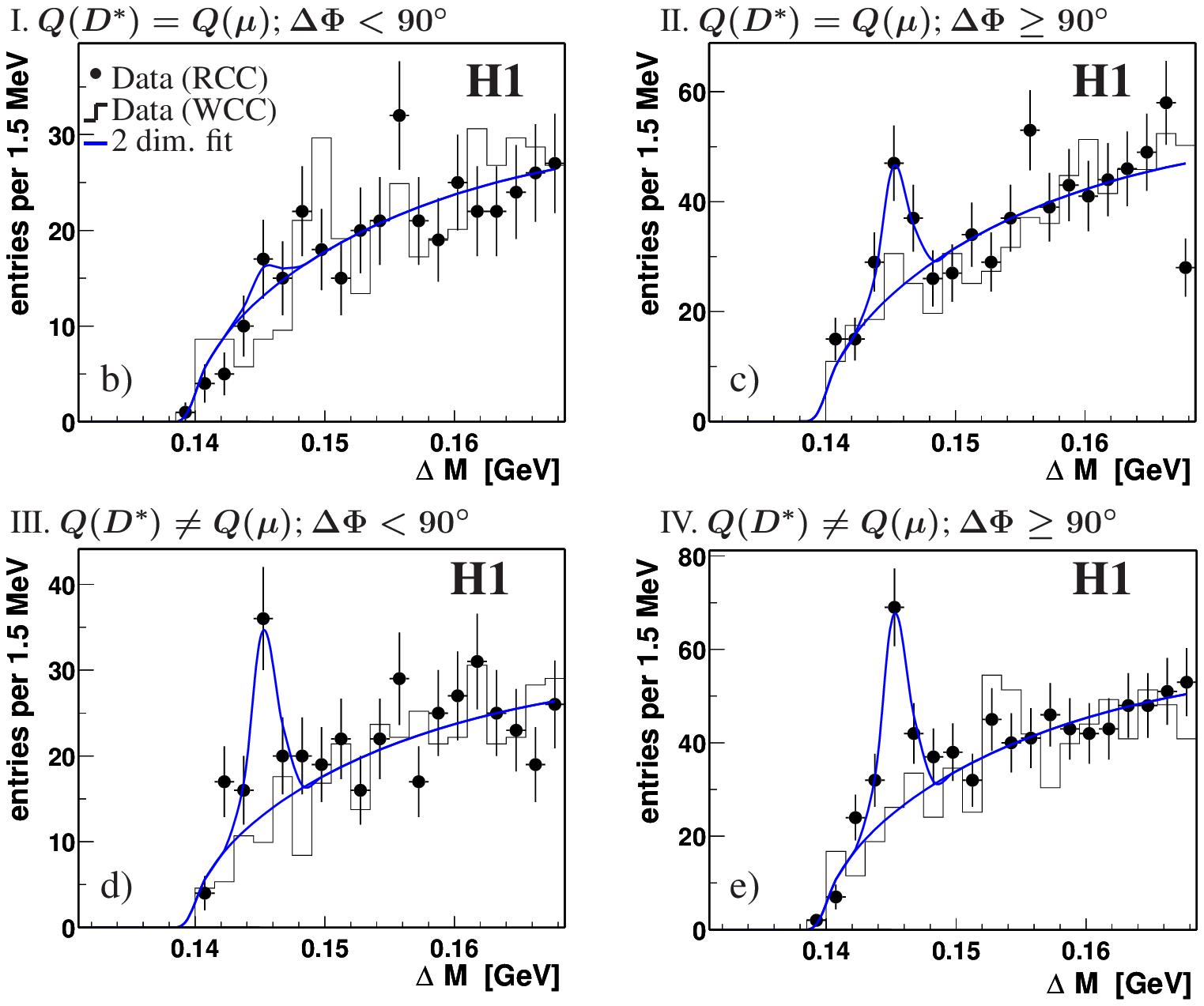,width=16.2cm,angle=0}}
 \end{tabular}
\vspace*{-1.2cm}
\caption{{\label{DsMufig7}}\em a) Distribution of the mass difference $\Delta M =
    m_{K\pi\pi_{s}}-m_{K\pi}$ for the total data sample. In b)-e) $\Delta M$ is 
    shown in the four
    correlation regions, given by the relative charges of the $D^{*}$ and the
    muon and the azimuthal angle $\Delta\Phi$ between them. 
    The points represent the data (right charge combinations, RCC), the histogram indicates the observed
    wrong charge combinations (WCC) which are also used to fit the background.
    The solid lines in a) are the result of a one-dimensional fit, which gives
    the peak position $\mu$ and the peak width
    $\sigma_{\Delta M}$. The solid lines in b)-e) are results of a
    two-dimensional fit (see text).}
  \end{center}
\end{figure}

The $D^{*}$ yield is measured using the $\Delta M$ technique~\cite{Feldman:1977ir},
where $\Delta M = m_{K\pi\pi_s}-m_{K\pi}$  is the difference of the invariant masses 
of the $K\pi\pi_s$ and the $K\pi$ systems.
Figure~\ref{DsMufig7}a shows the $\Delta M$ distribution for 
the selected $D^{*}\mu$ sample separately for the `right'
($K^{-}\pi^{+}\pi_{s}^{+}$) and `wrong' ($K^{-}\pi^{-}\pi_{s}^{+}$) charge combinations. 
The wrong charge distribution is normalised to the right charge distribution
in the range $0.155 \leq \Delta M  \leq 0.1685$~GeV.
The number of signal events is extracted from a
fit to the $\Delta M$ distribution using a Gaussian function for the signal and a 
parameterisation of the background\footnote{The functional form used is
                    $c_1\,(\Delta M-m_\pi)^{c_2}\,(1-c_3\,(\Delta M)^2)$ 
                    where $c_i$ are fit parameters.}. 
The parameters of the background function are determined from right and normalised wrong charge combinations.
The result of this fit for the total signal is also shown in figure~\ref{DsMufig7}a. 

The total number of \dstarmu\ events obtained from the fit is $N_{D^*\mu}=151\pm22$.
This number still contains a contribution from `fake muons', 
i.e. from hadrons  misidentified as muons and 
muons from the decay of light mesons. The fake muon background contributes
about 37\% in charm and 5\% in beauty initiated events according to the 
respective PYTHIA Monte 
Carlo simulations, which give an adequate description of the data.  

In figure~\ref{DsMufig7}b--e, the $\Delta M$ distributions of the selected 
\dstarmu\ events are shown separately for the four correlation 
regions defined in section \ref{corsec}.
Clear peaks due to \dstar\ mesons are observed in regions II-IV, 
whereas region I shows little or no signal, consistent with the expectation. 

The charm and beauty contributions in the data are determined by performing 
a simultaneous likelihood fit of the $\Delta M$ distributions 
 in the four correlation regions. 
In the following, this fit will be referred to as a 
`two-dimensional fit', in order to distinguish the results from the separate 
one-dimensional fits of $\Delta M$ in each correlation region. 

In this two-dimensional fit, in addition to the \dstarmu\ contribution from $b\bar{b}$ and
$c\bar{c}$, the fake muon background and the combinatorial background under the
$\Delta M$ peaks have to be considered.
The position and width of the $\Delta M$ peak corresponding to the \dstar\ signal as well as the parameters describing the shape of the
combinatorial background are fixed to the values obtained from the one-dimensional $\Delta M$ fit to the 
total sample (figure~\ref{DsMufig7}a). The normalisation of the combinatorial 
background is fitted using right and wrong charge combinations in each region separately.
The relative distributions of signal events from charm and beauty between 
the correlation regions as well as the fractions of fake muon background in each region predicted by
the PYTHIA Monte Carlo simulations are used as input for the fit.
In total there are six free fit parameters, the total numbers of \dstarmu\
events from  $b\bar{b}$ and $c\bar{c}$ quark pairs, $N_b$ and $N_c$, and 
four parameters for the combinatorial background, one in each correlation region. 

\section{Results} 
\begin{figure}
  \begin{center}
\setlength{\unitlength}{1.0cm}
\epsfig{file=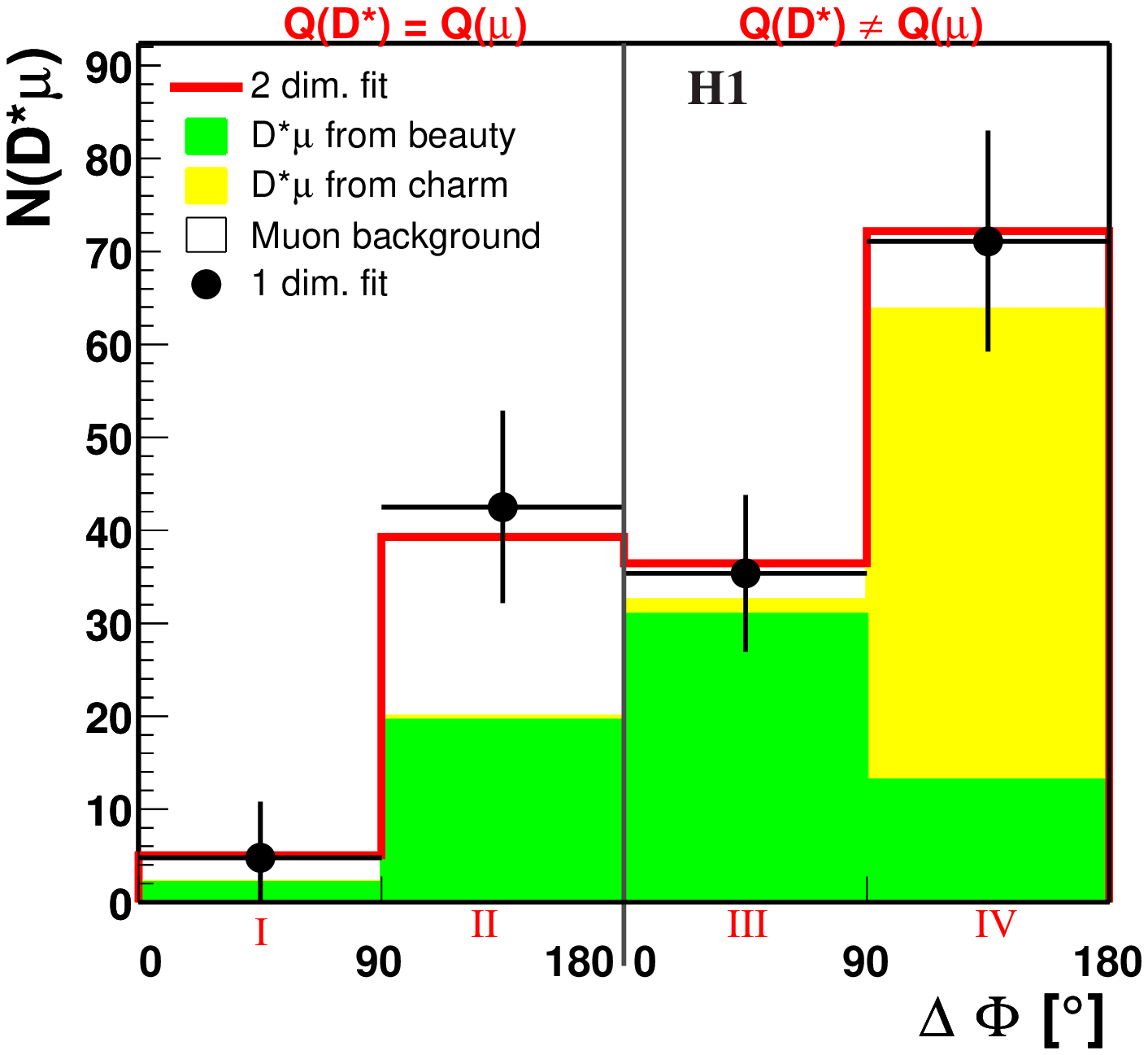,width=14cm}
\vspace*{-0.5cm}
\caption
{\em Population of the four correlation
regions I-IV obtained from the simultaneous likelihood fit in all correlation
regions (two-dimensional fit, histogram). The resulting decomposition into 
charm and beauty 
contributions and the muon background is also shown.
In all correlation regions the muon background is dominated by charm initiated events.
The points with error bars are the results 
of one-dimensional fits of the $\Delta M$ distributions in each correlation region.}
\label{DsMufig8}
  \end{center}
\end{figure}

The result of the two-dimensional fit 
is shown together with the data in figures~\ref{DsMufig7}b--e. 
The data are described well and the quality of the fit is good ($\chi^2=145.5$ for $154$ $d.o.f.$).
In figure~\ref{DsMufig8}, the numbers of \dstar\ signal events from the two-dimensional fit in 
the four correlation regions are compared to the results of one-dimensional fits of 
the $\Delta M$ distributions performed in each correlation region separately. The 
agreement is very good. The distribution of the contributing processes as obtained from the 
two-dimensional fit is also shown in figure~\ref{DsMufig8}.
The following event numbers and errors are 
obtained for the charm and beauty contributions from the two-dimensional fit: 
\begin{center}$N_c=53\pm12$\hspace{2cm}$N_b=66\pm17.$
\end{center}

\subsection{Total Cross Section}\label{secres}
The number of $b$ and $c$ events are used to compute the
total cross sections in the kinematic region defined in table~\ref{DsMuCuts}. 
The efficiencies and acceptances are derived from the 
Monte Carlo simulations. Values of 

$$\sigma_{vis}^{c}(ep\rightarrow e\,D^{*}\mu X)=250\pm 57\;(\mbox{stat}.)\pm 40\;(\mbox{syst}.)\;\mbox{pb}$$ 
and of 
$$\sigma_{vis}^{b}(ep\rightarrow e\,D^{*}\mu X)= 206\pm 53\;(\mbox{stat}.)\pm 35\;(\mbox{syst}.)\;\mbox{pb}$$

\noindent
are obtained for charm and beauty production, respectively. 
The measured cross sections are similar due to the definition
of the visible kinematic region, which requires in particular a high 
momentum muon, suppressing central charm production.
The results are compared with the pQCD predictions in table~\ref{totalCrossData1}, where 
error estimates due to the uncertainty of the quark masses
and the scales are given for the NLO calculations. 
In order to assess the influence of mass effects in the extraction of gluon
densities used in the calculations, the default CTEQ5M sets have been replaced by
the CTEQ5F sets~\cite{cteq5l}. The
results are found to be compatible. The uncertainties for PYTHIA and CASCADE are 
found to be of similar size as those of the NLO calculations. 
The measured cross section for charm production agrees 
well with the LO-ME+PS models (PYTHIA and CASCADE) and the NLO prediction (FMNR). 
The measured beauty cross section exceeds the calculated cross sections.
In other recent measurements of the beauty cross section in 
photoproduction at HERA~\cite{ourbtomupaper,Chekanov:2003}, 
based on the selection of high transverse momentum jets, 
ratios of measurement and FMNR based calculations between 
$1$ and $3$ are found. Note that the present analysis extends down to the production 
threshold for $b\bar{b}$, while the jet measurements have a threshold which is 
approximately $5$ GeV higher in the  $b\bar{b}$ centre-of-mass system.

\begin{table}[tb]
\begin{center}
\begin{tabular}{|l||c|c|}\hline
  {\textbf{Charm}} & Cross section [pb] & Data/Theory \\ \hline
\hline
 Data & $250\pm 57\pm 40$ & \\
 PYTHIA (direct) & $242~(142)$ & $1.0$ \\ 
 CASCADE & $253$ & $1.0$ \\ 
 FMNR & $286^{+159}_{-59}$ & $0.9$\\ \hline
 {\textbf{Beauty}}&&\\ \hline
 Data & $206\pm 53\pm 35$ & \\
 PYTHIA (direct) & $57~(44)$ & $3.6$ \\
 CASCADE & $56$ & $3.7$ \\ 
 FMNR & $52^{+14}_{-9}$ & $4.0$ \\ \hline
 \end{tabular}
\caption
{\em Measured $D^*\mu$ cross sections for charm and beauty production in the kinematic 
region defined in table~\ref{DsMuCuts}. For the data the statistical  
and the systematic errors are given. The LO-ME+PS predictions (PYTHIA, CASCADE) and  
NLO calculations (FMNR) are also shown. The uncertainties of the FMNR results 
are obtained by varying the renormalisation 
and the factorisation scales simultaneously by factors of 0.5 and 2. 
The uncertainty due to a variation of the quark masses  $m_c$ by $\pm 0.2$ GeV and 
$m_b$ by $\pm 0.25$ GeV is added quadratically. The last column shows the ratios of the measurement to the prediction.}
\label{totalCrossData1}
\end{center}
\end{table}

The systematic uncertainties of the cross section measurement are evaluated
by varying the Monte Carlo simulations.
The dominant experimental errors come from the uncertainties in the 
track reconstruction efficiency (13\%), the trigger efficiency (5\%) and the width of the
$\Delta M$ signal (3\%).
Smaller contributions are due to uncertainties in the  
determination of the background due to misidentified muons\footnote{Where two 
numbers are given the first applies to charm and the second to beauty.} 
(1\%;1.5\%) and in the fragmentation
fractions (1\%;1.5\%). Model uncertainties are 
estimated using the CASCADE Monte Carlo generator 
instead of PYTHIA (3.5\%) and either taking into account or omitting the resolved component in 
PYTHIA (3\%;5\%). Taking into account the uncertainties due to the contribution of
\dstar\ reflections (5\%), the muon identification, the luminosity
measurement and  the \dstar\ decay branching ratios, the total systematic errors
for the charm and beauty cross sections are estimated to be 16\% and 17\%, respectively.

\subsection{Differential Cross Sections for Charm and Beauty}
\begin{figure}
  \begin{center}
\setlength{\unitlength}{1.0cm}
\begin{picture}(16,17.5)
\put(0.,9.){\epsfig{file=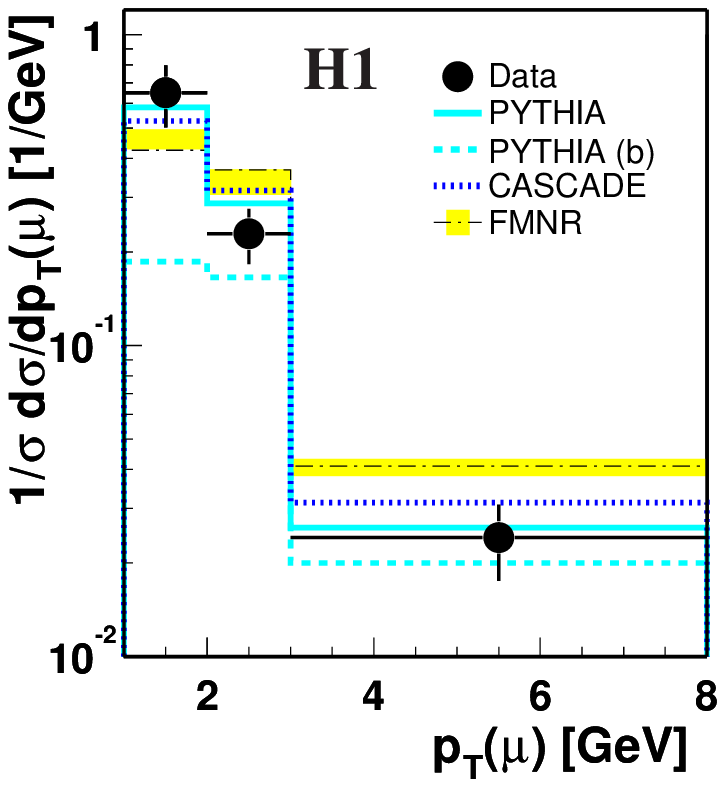,width=7.2cm}}
\put(8.,9.){\epsfig{file=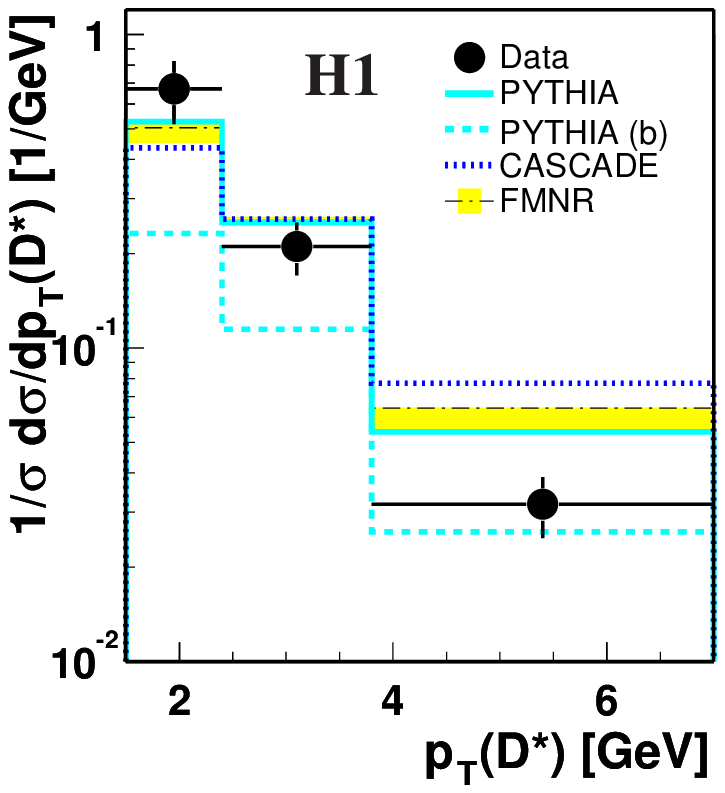,width=7.2cm}}
\put(0.,0.){\epsfig{file=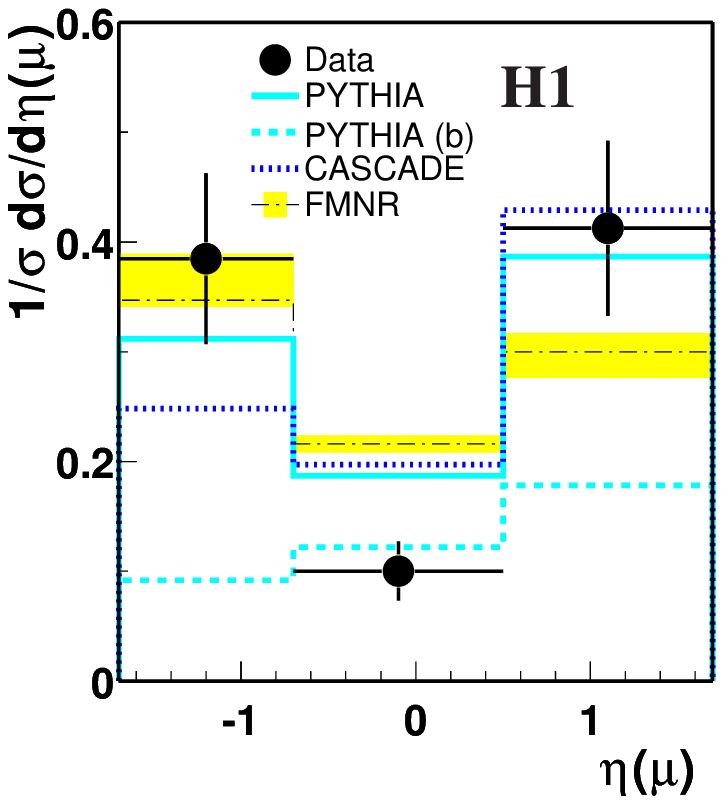,width=7.2cm}}
\put(8.,0.){\epsfig{file=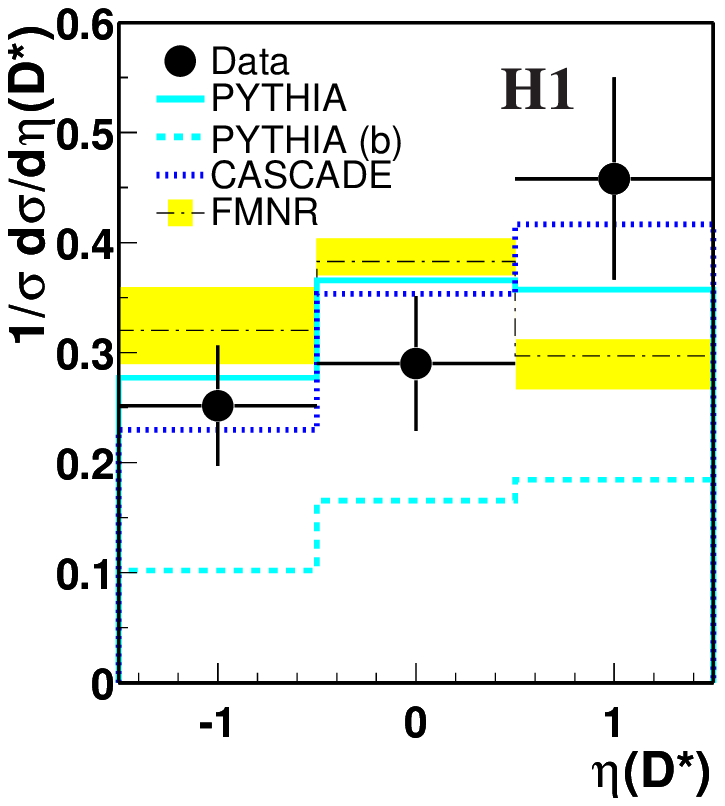,width=7.2cm}}
\put(5.,17.3){\Large CHARM AND BEAUTY }
\put(1.5,10.7){\large a) }
\put(9.5,10.7){\large b) }
\put(1.5,1.6){\large c) }
\put(9.5,1.6){\large d) }
\end{picture}
\caption{\em Normalised differential $D^*\mu$ cross sections as
  functions of the transverse momenta and the pseudorapidities 
of muons (a,c) and \dstar\ mesons (b,d). The data (points) are compared with the 
prediction of the NLO calculation FMNR and the LO-ME+PS QCD models
PYTHIA and CASCADE. A beauty fraction of 45\% as obtained from the measured cross
sections is used in the calculations. The error bands for FMNR are obtained as
 described in table \ref{totalCrossData1}. The PYTHIA $b$ quark contribution is indicated separately.
The experimental systematic uncertainties for the normalised distributions are negligible compared 
to the statistical errors.}
\label{SigmaGP2}
\end{center}
\end{figure}

\begin{figure}
  \begin{center}
\setlength{\unitlength}{1.0cm}

\begin{tabular}{cc}
\hspace*{-0.5cm}\epsfig{file=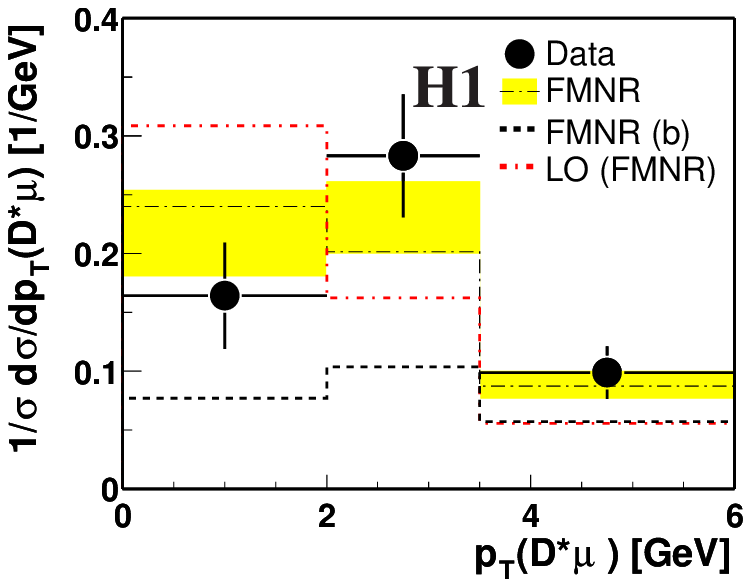,width=7.4cm,clip=}
& \hspace*{0.2cm}\epsfig{file=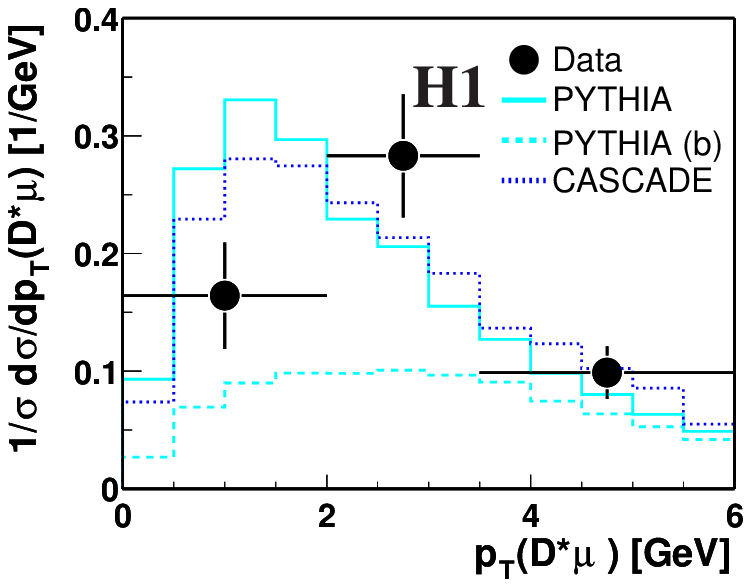,width=7.4cm,clip=} \\
\hspace*{-0.5cm}\epsfig{file=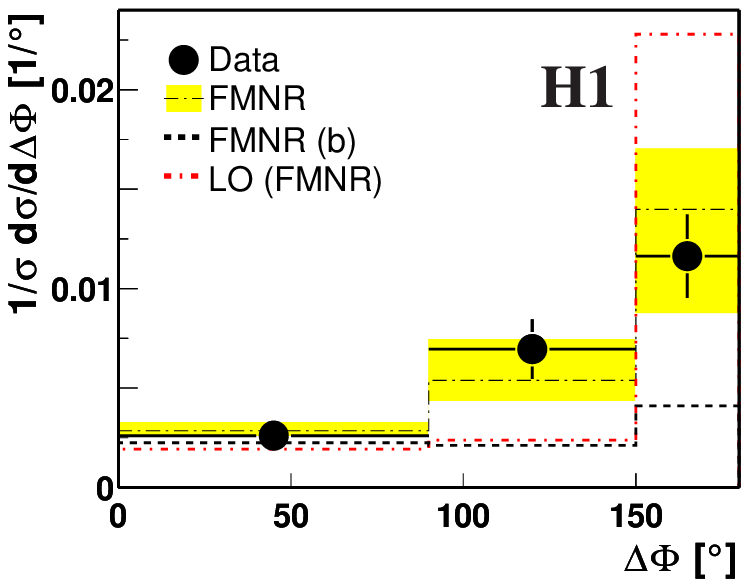,width=7.4cm,clip=}
& \hspace*{0.2cm}\epsfig{file=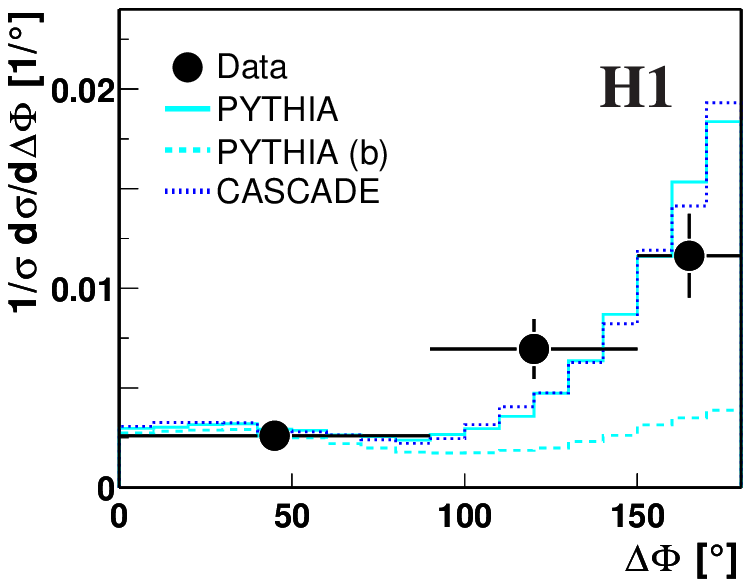,width=7.4cm,clip=} \\
\hspace*{-0.4cm}\epsfig{file=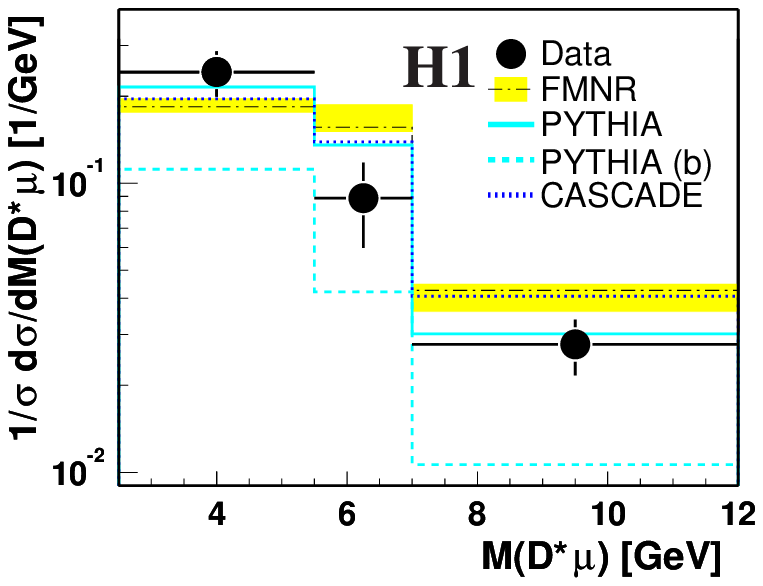,width=7.4cm,clip=}
 & \hspace*{0.3cm}\epsfig{file=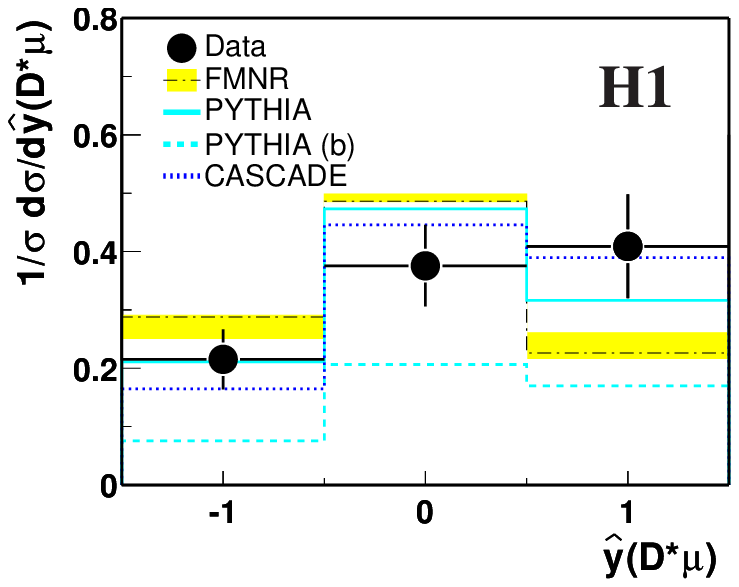,width=7.55cm,clip=}
\end{tabular}
\begin{picture}(16,2.1)
\put(5.,20.2){\Large CHARM AND BEAUTY }
\put(1.5,15.4){\large a) }
\put(9.9,15.4){\large b) }
\put(1.5,10.5){\large c) }
\put(9.6,10.5){\large d) }
\put(1.5,3.5){\large e) }
\put(9.6,3.45){\large f) }
\end{picture}
\vspace*{-2.8cm}
\caption{\em Normalised differential $D^*\mu$ cross sections for 
a,b) the transverse momenta $p_T(D^*\mu)$, 
c,d) the azimuthal angle difference $\Delta\Phi$, e) the invariant mass $M(D^*\mu)$ and f) 
the rapidity 
$\hat{y}(D^*\mu)$ of the $D^*\mu$-pairs. The data are compared to the 
prediction of the LO and NLO calculations FMNR (a,c,e,f) and to the 
Monte Carlo models PYTHIA and CASCADE (b,d,e,f). The error bands for FMNR are obtained as
described in table \ref{totalCrossData1}. A beauty fraction of 45\%
as obtained from the measured cross sections is used in the calculations. The
FMNR (a,c) and PYTHIA (b,d,e,f) $b$ quark 
contributions are indicated separately. The experimental systematic uncertainties for the normalised distributions are negligible compared 
to the statistical errors.}
\label{SigmaGP1}
\end{center}
\end{figure}

Differential cross sections for $D^{*}\mu$ production in the visible 
kinematic region  are evaluated 
as functions of variables characterising the \dstar\ meson, the muon and the \dstarmu\ system.
In this section results are presented for the complete data set, 
which contains the contributions from charm and beauty (figures~\ref{SigmaGP2} and \ref{SigmaGP1}).  

In order to compute the differential cross sections for the data, the numbers of 
events in bins of the chosen variable are determined by a 
fit to the $\Delta M$ distribution in each bin, as described in section
\ref{secfit}. Here, no attempt is made to separate charm and beauty contributions.
A correction for 
`fake muons' is applied according to the Monte Carlo simulation. Since the fake muon fraction 
is different for charm and beauty, it is computed using the $b$ 
fraction of 45\% given by the measured cross sections (table~\ref{totalCrossData1}). 

The data are shown with the results of the PYTHIA and CASCADE Monte Carlo models 
and the LO and NLO FMNR calculations. In the theoretical 
models, the beauty and charm contributions are combined according to the
measured total visible cross sections (table~\ref{totalCrossData1}) and normalised to
the sum of these cross sections, in order to facilitate a shape comparison.
The error bands for the FMNR prediction are computed as for the total cross section (see caption of table~
\ref{totalCrossData1}). The measured differential cross sections are similarly
normalised, which 
has the advantage that the systematic errors largely
cancel and are negligible compared to the statistical errors.

Figure~\ref{SigmaGP2} shows the differential cross sections as a function of
the transverse momentum and pseudorapidity of the \dstar\ meson and the muon separately. 
Overall the QCD models  describe the shapes of the measured distributions quite well, 
although there is a tendency for the measured $p_{T}(D^{*})$ and $p_{T}(\mu)$
distributions to be softer than the calculations. A slight discrepancy is also
present in the differential cross section as a function of the pseudorapidity
of the muon (figure~\ref{SigmaGP2}c) which shows a central dip due to the large muon momentum required.

Quantities derived from a combined measurement of the \dstar\ and muon are
shown in figure~\ref{SigmaGP1}. In figures~\ref{SigmaGP1}a and c, the 
differential cross sections as a function of  $p_{T}(D^{*}\mu)$, which is 
defined as $p_{T}(D^{*}\mu)=|\vec{p}_{T}(D^{*})+\vec{p}_{T}(\mu)|$, 
and $\Delta \Phi$ are compared with the LO and NLO FMNR predictions.
The data show the expected deviations from the LO calculations due to higher order effects:
the observed $p_{T}(D^{*}\mu)$ distribution is flatter and the $\Delta \Phi$ peak 
around $180^\circ$ is broader than the LO
computation. The data are in good agreement with the NLO calculation. 
In figures \ref{SigmaGP1}b and d, the same differential cross sections for 
$p_{T}(D^{*}\mu)$ and $\Delta \Phi$ are compared with PYTHIA and CASCADE
which also give a good description of the data. Although different approaches 
are used in PYTHIA and CASCADE to compute the evolution of the partons from the proton 
and the hard interaction, the differences between the two simulations
are smaller than the experimental errors.

Figures \ref{SigmaGP1}e and f show the invariant mass, $M(D^{*}\mu)$, and the 
rapidity\footnote{$\hat{y} = 1/2 \, \ln{(E+p_z)/(E-p_z)}$, 
where $E$ and $p_z$ are the energy and the $z$-component of the momentum of the \dstarmu\ pair.}, 
$\hat{y}(D^{*}\mu)$,
of the \dstar\ meson and the muon together with NLO FMNR, PYTHIA and CASCADE predictions. 
The invariant mass $M(D^{*}\mu)$ reflects the centre-of-mass energy of the quark pair 
and $\hat{y}(D^{*}\mu)$ is related to the ratio of the energies of the partons entering
the hard interaction from the proton and the photon.
Both differential cross sections are adequately described by all model calculations.

\subsection{Results for a Charm Dominated `Quark Antiquark Tag' Sample}\label{2qtag}
The cross sections in the previous section refer to the complete data set
including events from region III in which both \dstar\ and muon 
originate from the same $b$ quark.
Since this leads to a dilution of the correlation of
quantities characterising the quark pair and the measured \dstarmu\ pair,  
results for a smaller sample are given here, where both the heavy quark and the
antiquark are tagged by either a \dstar\ or a muon 
(`quark antiquark tag').  This is possible in correlation region IV 
($\Delta\Phi> 90^{\circ}$ and $Q(D^{*})\neq Q(\mu)$).
This region (see figure~\ref{sketch}) 
is dominated by $c\bar{c}$ pairs: the 
$b\bar{b}$ contribution is 18\% according to the two-dimensional fit. 
Due to migrations from correlation region III, approximately half of the 
$b\bar{b}$ contribution is due to $b\rightarrow D^*\mu$ events in which the 
\dstarmu\ pair comes from the same $b$ quark (according to the PYTHIA 
simulation). 
A visible cross section of  
$\sigma_{2q}=263\pm48\pm36$~pb is measured\footnote{The index `$2q$' is used
for the cross sections in this section to distinguish them from those in the
previous section.} after subtracting this fraction, while $264^{+148}_{-50}$~pb is 
expected from the FMNR calculations.

\begin{figure}
 \setlength{\unitlength}{1.0cm}
 \begin{center}
 \begin{tabular}{cc}
\epsfig{file=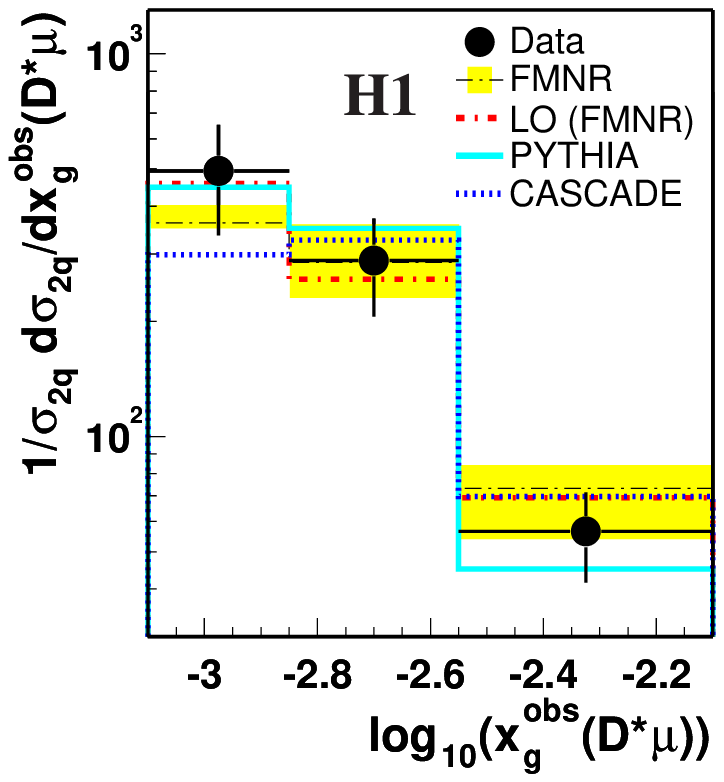,width=7.2cm}
 &
\epsfig{file=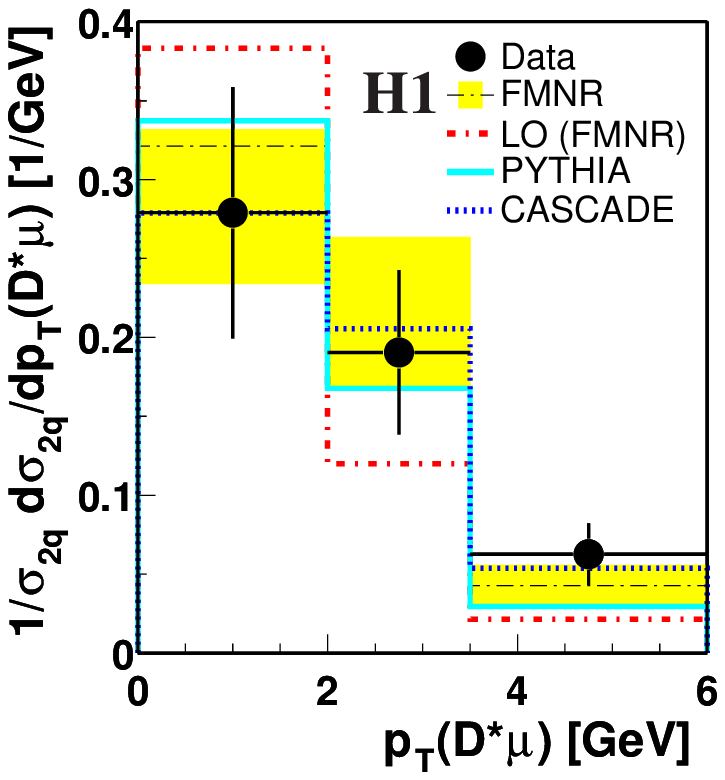,width=7.2cm}
 \end{tabular}

\vspace*{0.5cm}\epsfig{file=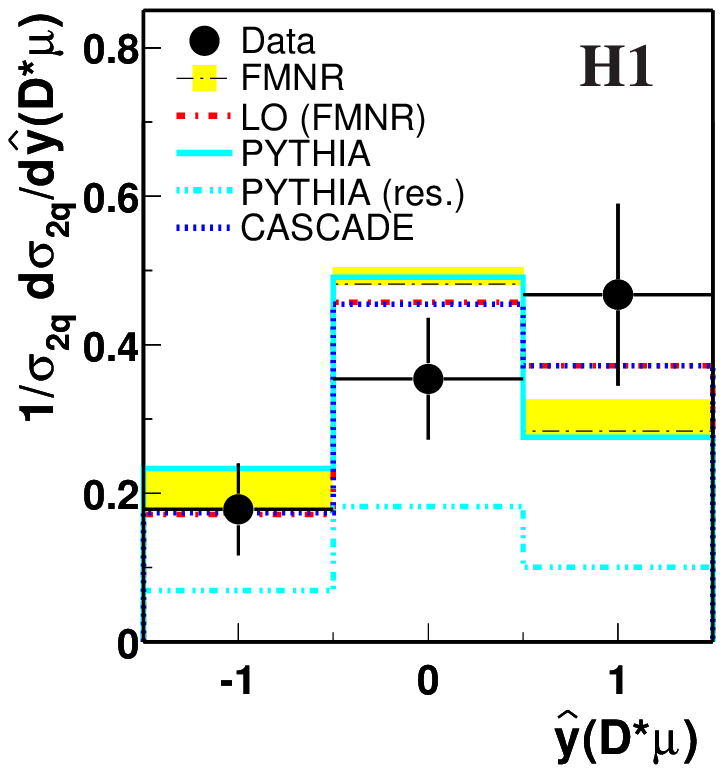,width=7.2cm,clip=}
\begin{picture}(16,2)
\put(4.5,19.){\Large QUARK ANTIQUARK TAG}
\put(2.2,12.){\large a) }
\put(9.7,12.){\large b) }
\put(8.1,3.5){\large c) }
\end{picture}
\vspace*{-2.5cm}\caption{\em Normalised differential $D^*\mu$ cross sections for a
`quark antiquark tag', charm dominated sample (approximately 10\% $b\bar{b}$ quark contamination), 
where the $D^*$ and the $\mu$ originate from different quarks.
The data and predictions of the LO and NLO calculation FMNR and of the 
Monte Carlo generators PYTHIA and CASCADE are shown.  The error bands for FMNR are obtained as
described in table \ref{totalCrossData1}. In c) the resolved excitation component of 
PYTHIA is indicated separately. The experimental systematic uncertainties for the normalised 
distributions are negligible compared to the statistical errors.}
\label{SigmaGP3}
\end{center}
\end{figure}

In this data sample, the correlation of kinematic quantities reconstructed
using the \dstar\ meson and the muon to those of the quark pairs is good for $x_g$ and 
$\hat{y}(D^{*}\mu)$, while it is weaker for $p_T($\dstarmu). Here
$x_g$ is the fraction of the proton energy carried by the gluon in the hard
interaction, which is approximated by $x^{obs}_g=(M(D^{*}\mu))^2/y\,s$.
The normalised differential cross section for $x^{obs}_g(D^*\mu)$ is shown in 
figure~\ref{SigmaGP3}a. 
All QCD calculations (FMNR to LO and NLO, PYTHIA and CASCADE) give a reasonable 
description of the data.
Figure~\ref{SigmaGP3}b and c show the $p_T($\dstarmu)\ and  
$\hat{y}(D^{*}\mu)$ distributions of the \dstarmu\ pair, respectively, with the 
same model calculations. The LO FMNR prediction for $p_T($\dstarmu) is again too soft, 
as observed for the total sample (figure~\ref{SigmaGP1}a), while the NLO FMNR 
prediction fits the data well.
Although this sample should be sensitive to any transverse momentum of the incoming gluon, the
differences between PYTHIA (collinear factorisation) and CASCADE 
($k_t$ factorisation) are small in the kinematic region studied.

\section*{Conclusion}
A measurement of $c\bar{c}$ and $b\bar{b}$ photoproduction cross sections using the H1 detector at 
HERA has been presented. For the majority of events both heavy quarks are tagged  
using a $D^{*}$ meson and a muon as signatures.
The separation of the charm and beauty contributions is possible due to the different
correlations between the charges and angles between the \dstar\ meson and the muon.
The measured total cross section for charm in the visible kinematic region 
is in agreement with the NLO QCD prediction, while the beauty cross section is higher
than predicted.
The kinematic region of the latter is characterised 
by lower $b\bar{b}$ centre-of-mass energies than in most previous 
analyses, which require high momentum jets. 
Comparisons of the shapes of the measured differential distributions with QCD 
calculations including higher order effects show general agreement.
Effects beyond the LO approximation are directly observed.
In the kinematic region studied, effects due to $k_t$ factorisation, as
implemented in CASCADE, are found to be small compared to the experimental
errors. 
\section*{Acknowledgements}
We are grateful to the HERA machine group whose outstanding
efforts have made this experiment possible. 
We thank the engineers and technicians for their work in constructing and
maintaining the H1 detector, our funding agencies for 
financial support, the DESY technical staff for continual assistance
and the DESY directorate for support and for the hospitality which they extend to the non DESY 
members of the collaboration.

\newpage%

\clearpage



\end{document}